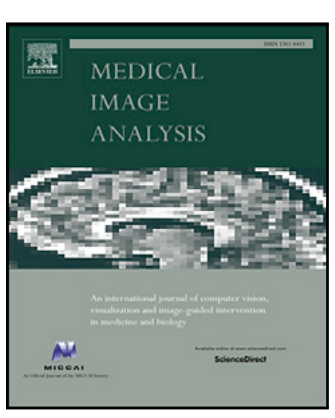

# Towards cardiac MRI foundation models: Comprehensive visual-tabular representations for whole-heart assessment and beyond

Yundi Zhang [a,b,*], Paul Hager [a,b], Che Liu [c], Suprosanna Shit [f], Chen Chen [d,e,1], Daniel Rueckert [a,b,d,1], Jiazhen Pan [a,b,1]

[a] School of Computation, Information and Technology, Technical University of Munich, Munich, Germany
[b] School of Medicine, Klinikum Rechts der Isar, Technical University of Munich, Munich, Germany
[c] Data Science Institute, Imperial College London, London, UK
[d] Department of Computing, Imperial College London, London, UK
[e] Department of Computer Science, University of Sheffield, Sheffield, UK
[f] Department of Quantitative Biomedicine, University of Zurich, Zurich, Switzerland

A B S T R A C T

Cardiac magnetic resonance (CMR) imaging is the gold standard for non-invasive cardiac assessment, offering rich spatio-temporal views of the heart's anatomy and physiology. Patient-level health factors, such as demographics, metabolic, and lifestyle, are known to substantially influence cardiovascular health and disease risk, yet remain uncaptured by CMR alone. To holistically understand cardiac health and to enable the best possible interpretation of an individual's disease risk, CMR and patient-level factors must be jointly exploited within an integrated framework. Recent multi-modal approaches have begun to bridge this gap, yet they often rely on limited spatio-temporal data and focus on isolated clinical tasks, thereby hindering the development of a comprehensive representation for cardiac/health evaluation.

To overcome these limitations, we introduce *ViTa*, a step toward foundation models that delivers a comprehensive representation of the heart and a precise interpretation of individual disease risk. Leveraging data from 42,000 UK Biobank participants, ViTa integrates 3D+T cine stacks from short-axis and long-axis views, enabling a complete capture of the cardiac cycle. These imaging data are then fused with detailed tabular patient-level factors, enabling context-aware insights. This multi-modal paradigm supports a wide spectrum of downstream tasks, including cardiac phenotype and physiological feature prediction, segmentation, and classification of cardiac/metabolic diseases within a single unified framework. By learning a shared latent representation that bridges rich imaging features and patient context, ViTa moves beyond traditional, task-specific models toward a universal, patient-specific understanding of cardiac health, highlighting its potential to advance clinical utility and scalability in cardiac analysis. [2]

## 1. Introduction

Cardiac magnetic resonance (CMR) imaging is the gold standard for non-invasive cardiac assessment, providing detailed insights into heart anatomy, function, and morphology (Myerson, 2021). However, while CMR excels at capturing the heart's anatomy and function, broader information is embedded in patient-level health factors, such as sex, BMI, and lifestyle, that significantly influence cardiac health (Zhu et al., 2022). To achieve true personalized cardiac healthcare, a comprehensive, holistic understanding of an individual's cardiac state, incorporating both structural insights and patient-specific context, is urgently needed. However, the progress in developing holistic representations for cardiac models has been lacking in three different aspects: (1) integration of full sequence CMR data, (2) integration of broader patient health factors, and (3) a latent representation foundation that can simultaneously support a wide range of downstream tasks.

*Lack of incorporation of entire spatio-temporal CMR data:* A standard CMR acquisition consists of cine sequences from multiple views (Petersen et al., 2015): several parallel short-axis (SA) views and 3 long-axis (LA) views oriented perpendicularly to the SA slices. This setup enables efficient 3D+time (3D+T) imaging of the heart, capturing the entire cardiac cycle and providing a uniquely powerful means of assessing dynamic cardiac function. However, current CMR-based models often do not leverage this full spatio-temporal richness for cardiac analysis (Chen et al., 2020b). Due to high computational costs, many models reduce full-cycle CMR data to a few manually selected views or focus on only the end-diastole and end-systole frames, assuming these are sufficient to characterize cardiac function. However, these approaches ignore the continuous nature of cardiac structure and motion, not only hindering clinical decision-making but also preventing the development of a universal, comprehensive understanding of cardiac health.

* Corresponding author.
*E-mail addresses:* yundi.zhang@tum.de (Y. Zhang), jiazhen.pan@tum.de (J. Pan).
[1] These authors advice equally.
[2] The code is open-sourced at https://github.com/Yundi-Zhang/ViTa.git.

*Lack of integration of broader patient health context:* In real-world clinical practice, clinicians rely on a holistic understanding of the heart that combines diverse sources of information, such as structural and functional parameters of the cardiovascular system as well as systemic health parameters. While CMR imaging provides detailed structural and functional insights, it alone cannot form a holistic view of a patient's cardiac health. Key factors such as demographics, metabolic status, and lifestyle habits are essential for interpreting cardiac function in a patient-specific manner. However, most existing models rely solely on imaging data (Jacob et al., 2024; Kim et al., 2024), ignoring these systemic influences and failing to incorporate contextual patient information into the assessment of cardiac health. This oversight leads the model to focus on image-level information, lacking a holistic understanding of the heart. As a result, current models fail to advance toward personalized, context-aware cardiac health evaluation.

*Lack of a comprehensive representation foundation for versatile downstream tasks:* Current approaches to cardiac analysis tend to focus on specific clinical tasks, such as segmentation (Schilling et al., 2024), phenotype prediction (Lange et al., 2024), or disease classification (Ma et al., 2024), each requiring a separately trained model. This fragmented training strategy prevents the development of a unified, generalizable representation of cardiac function and demands task-specific retraining for every new application. As a result, models struggle to generalize across tasks and fail to capture a holistic understanding of cardiac health.

***Our Contributions.*** To overcome these three limitations, we introduce *ViTa*, a comprehensive cardiac representation that integrates complete 3D+T cardiac information with patient-level health data for holistic and versatile cardiac analysis. Specifically,

- We introduce *ViTa*, a fully multi-view, multi-modal, multi-task model for comprehensive cardiac function assessment, enabling a dynamic, 4D understanding of the heart. Furthermore, this unified model seamlessly integrates phenotype and physiological feature prediction, cardiac segmentation, and cardiac/metabolic disease classification with imbalanced classes into a single framework. Our work makes an initial step toward a comprehensive cardiac function/overall health assessment that could be deployed for multiple applications and tasks.
- We generate comprehensive and information-rich whole-heart representations by integrating both CMR images and tabular health factors, enabling large-scale multi-modal cardiac analysis. By aligning imaging-derived cardiac function with patient-specific demographic, metabolic, and lifestyle factors, our model learns intrinsic, physiology-driven representations that bridge cardiac structure, function, and systemic health. This approach moves beyond traditional imaging biomarkers and enhances the understanding of cardiac health with patient context, paving the way toward a foundation model for holistic, personalized cardiac care.
- We demonstrate that patient-specific cardiac analysis enables the capture of functional disease patterns conditioned on individual physiological and lifestyle factors, leading to accurate cardiac phenotype and physiological feature prediction with all features derived simultaneously, as well as superior classification of cardiovascular and metabolic diseases with imbalanced classes. This personalized approach enhances diagnostic accuracy, allowing *ViTa* to learn beyond image-level information and uncover clinically relevant, individualized disease interpretation.

## 2. Related work

### 2.1. Multi-view and multi-frame CMR analysis

Despite the potential of full 3D+T CMR sequences, relatively few studies have fully leveraged multi-view and multi-frame data for comprehensive cardiac analysis. While various methods have been developed for cardiac segmentation (Petitjean and Dacher, 2011; Bai et al., 2017; Chen et al., 2020b; Campello et al., 2021; Zhang et al., 2024b), which provide a description of the cardiac anatomy and derive phenotypic indicators (Bai et al., 2020), these models often rely on sparse views of the heart. For instance, Qin et al. (2018) utilized spatio-temporal smoothness for segmentation but focused only on SA views, ignoring LA views. Similarly, Chen et al. (2019) incorporated multi-view anatomical priors for segmentation guidance but did not leverage the temporal information presented in cine CMR data. Stolt-Ansó et al. (2023) proposed a high-resolution 3D segmentation approach using neural implicit functions, yet this model omitted integration of LA views. Although cardiac motion tracking offers another avenue for understanding heart morphology (Wang et al., 2021; Meng et al., 2022; Pan et al., 2021, 2022, 2024a; Ghoul et al., 2024), these methods often do not fully capture the cardiac anatomy and dynamics from multi-view cine data, lacking a comprehensive 4D understanding of the cardiac dynamic function.

### 2.2. Multi-modal learning for cardiac assessment

While CMR images provide vital information on cardiac structure, function, and pathology, they only provide a partial perspective of the patient's cardiac health (Jafari et al., 2023). Recent advances in cardiac multi-modal learning have shown that combining CMR data with other complementary data sources, such as echocardiography (Puyol-Antón et al., 2022) and electrocardiograms (ECG) (Turgut et al., 2025; Radhakrishnan et al., 2023; Ding et al., 2024; Alsekait et al., 2024), enables a more comprehensive understanding of cardiovascular conditions.

Incorporating tabular data, which includes patient-level health factors like demographics, metabolic status, and lifestyle factors, has also gained significant attention due to its ability to provide high-level, informative features that enable effective end-to-end cardiac multi-modal learning (Borisov et al., 2022). For instance, Hager et al. (2023) fused tabular data with mid-ventricular 2D SA CMR for cardiac disease classification, while Du et al. (2024) developed it further by integrating incomplete tabular data with CMR data to enable robust cardiac disease classification with missing tabular values. However, they rely on a limited set of SA views, missing the opportunity to leverage the complete spatial and temporal context of CMR data. In addition, the absence of image encoder pretraining limits their capacity to learn rich, generalizable cardiac representations. Tripathi et al. (2024) integrated SA and 4-chamber CMR data together with electronic health records for pulmonary arterial wedge pressure prediction. However, their approach was also tailored to a specific task.

### 2.3. On the way to cardiac foundation models

Recent advances in biomedical foundation models have demonstrated strong potential for multi-task learning across diverse healthcare applications (Moor et al., 2023; Tu et al., 2024; Li et al., 2024; Zhang et al., 2024c; Zhou et al., 2024; Pan et al., 2024b; Liu et al., 2025). Despite growing interest in foundation models trained with diverse modalities, their application in multi-task cardiac assessment remains largely unexplored. Recent efforts have begun addressing this gap: Christensen et al. (2024) proposed a vision-language interpretation model fusing echocardiography and expert reports unified for cardiac phenotype prediction and patient identification, while Jacob et al. (2024) conducted a broad evaluation across 9 clinical tasks using large-scale CMR images with various modalities. Qiu et al. (2023) and Shad et al. (2023) focused on cardiomyopathy classification by integrating clinical reports with CINE/LGE multi-view imaging. Sørensen et al. (2024) leveraged Color Fundus Photography scans for spatio-temporal cardiac feature modeling guided by clinical demographics. Zhang et al. (2023) utilized contrastive learning to integrate 2D X-rays and single-slice CT/MRI data for image classification and cross-modal retrieval, while Curran et al. (2023) incorporated genetic information with CINE SA data for hypertrophic cardiomyopathy phenotype classification. Xia et al. (2022) embedded both CMR and demographic information for 3D patient-specific cardiac shape generation. However, the integration of whole CMR information with tabular clinical data for multi-task cardiac assessment remains an unexplored opportunity, offering significant potential for more personalized and comprehensive cardiac health analysis.

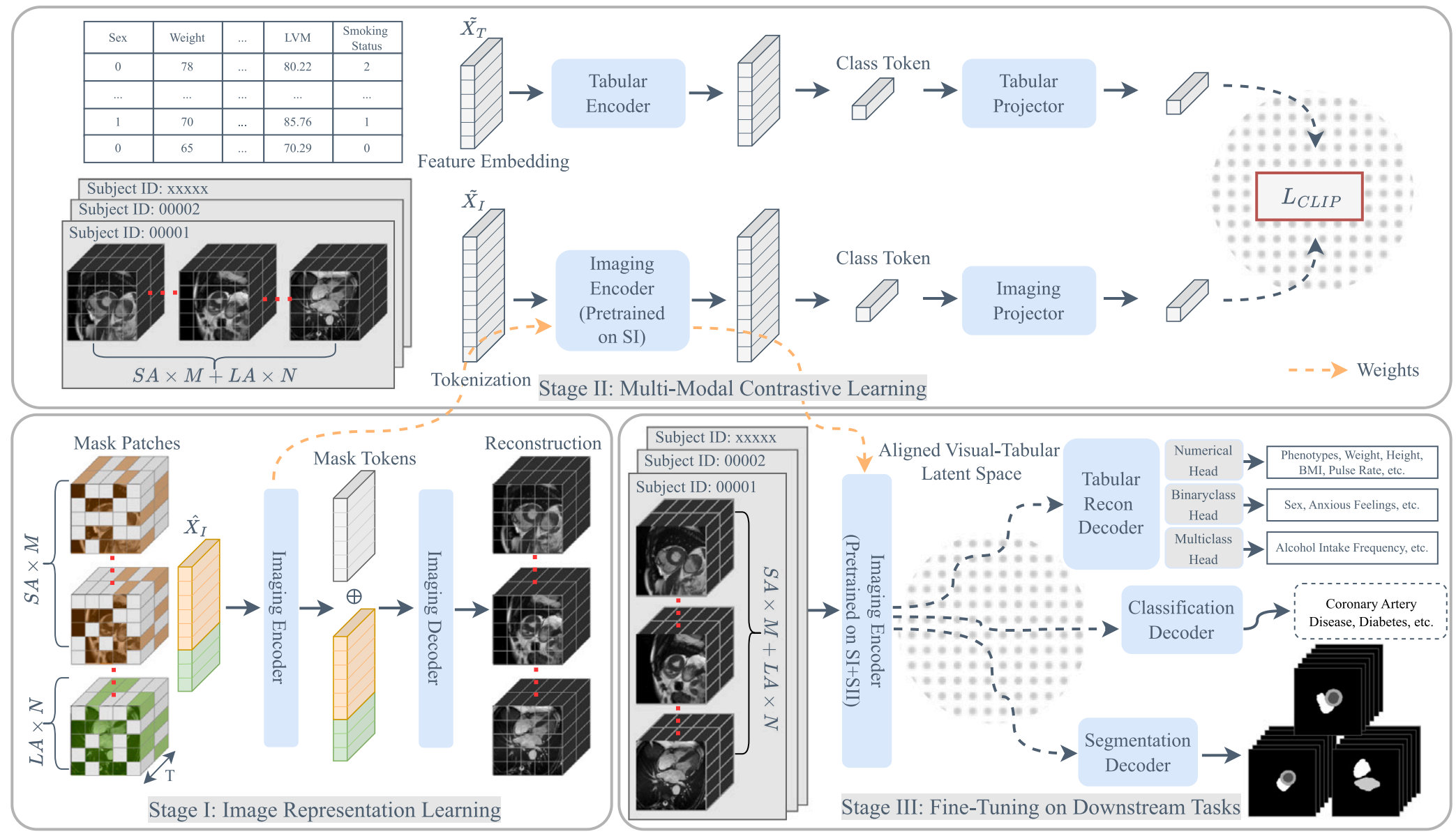

**Fig. 1.** Overview of *ViTa*. Stage I: Image representations are learned through self-supervised reconstruction of a stack of multi-view, multi-frame masked CMR slices (6 SA and 3 LA). Stage II: Image representations and tabular embeddings are aligned using contrastive learning by applying a CLIP loss. Stage III: Once aligned, the comprehensive cardiac representations are derived exclusively from CMR images and used for downstream tasks, including cardiac phenotypes and patient-level health factor prediction, disease classification, and all-plane segmentation. The imaging encoder is fine-tuned for downstream tasks.

## 3. ViTa

Our objective is to construct an enriched, dense, and comprehensive latent space that incorporates information from 3D+T CMR images and demographic/anthropometric data, providing a rich and versatile representation foundation that can be used for multiple downstream tasks in the assessment of cardiovascular health. To achieve this, we utilize contrastive learning (Chen et al., 2020a) due to its efficiency and strength in multi-modal feature alignment. However, the effectiveness of contrastive learning is contingent on large batch sizes, which is impractical with high-dimensional 3D+T CMR data. To address this, we train an imaging encoder through a multi-view MAE (Zhang et al., 2024a) approach, allowing us to generate compressed imaging embeddings for contrastive learning. ViTa's training process is carried out in three stages as illustrated in Fig. 1.

### 3.1. Stage I: Image representation learning

In the first stage, we utilize a Masked Autoencoder (MAE) (He et al., 2022; Zhang et al., 2024a) for self-supervised learning. The primary goal of this stage is to enable the imaging encoder to capture the underlying multi-view, dynamic features of the 3D+T CMR data. By masking portions of the image, we encourage the model to learn a robust and generalized representation of cardiac structures and dynamics across various views and frames, allowing it to understand the continuity of cardiac motion and structural variations. This pretraining stage effectively compresses the imaging data, enabling the imaging encoder to learn meaningful representations from high-dimensional CMR data, which facilitates the integration of multi-modal data in subsequent stages.

Each CMR scan contains sparse $M$ SA planes and $N$ LA planes, oriented perpendicularly to each other, forming a 3D+T view of the heart. Treating them as a unified 4D tensor is inefficient due to significant dissimilarities in voxel resolution and spatial orientation between views. Thus, we decompose each plane into 3D patches of shape $(p_x, p_y, p_t)$, and a 3D convolution with a stride equal to the patch size is applied to extract patch-level tokens that capture local structure. A random masking strategy is then applied, discarding $q\%$ of patches to reduce computational burden and promote robust feature learning. A high masking ratio encourages the encoder to capture spatio-temporal correlations across cardiac planes, rather than simply interpolating missing values. The remaining visible patches, denoted as $\hat{X}_I \in \mathcal{R}^{L \times dim}$ — where $L$ is the number of remaining patches and $dim$ is the feature dimension — are processed by an encoder $\mathcal{E}_I$ to learn a dense 3D+T cardiac representation. We extend positional embeddings (Vaswani et al., 2017) to capture 5D positional context: 3D for $x$-$y$-$t$, 1D for identifying the plane in the stack, and 1D for identifying the view (SA or LA) (for more details see Appendix B). Finally, a lightweight decoder $\mathcal{D}_I$ reconstructs the original input by reconstructing the original pixels of the masked patches, refining the learned CMR representations.

### 3.2. Stage II: Multi-modal embedding alignment with CLIP loss

In the second stage, we focus on integrating patient-level tabular health factors with the cardiac images. We achieve this by efficiently aligning the image embeddings (from the MAE encoder) with tabular embeddings.

Tabular data includes anthropometrics, phenotypes, physiological features, and lifestyle factors, with three different formats: numerical features, binary categorical features, and general categorical features. We first use two embedding strategies: linear-scaled embeddings, $\Theta_n$, for numerical features and learnable lookup table embeddings, $\Theta_c$, for both binary and general categorical features. We denote numerical features as $T_n$ and categorical features as $T_c$, which include both binary and general categorical features. For each subject, the initial tabular embeddings are formed by concatenating the embeddings of both feature types:

$$\tilde{X_T} = \Theta_n(T_n) \oplus \Theta_c(T_c), \tag{1}$$

where $\oplus$ denotes concatenation. A transformer-based encoder $\mathcal{E}_T$ is then applied to generate the final tabular embeddings: $\mathcal{E}_T(\tilde{X_T})$.

To ensure effective alignment, we utilize a contrastive loss function inspired by CLIP (Contrastive Language-Image Pretraining) (Chen et al., 2020a). This loss function pulls the learned embeddings from imaging and tabular data into a shared latent space, where similar patients (in terms of both cardiac features and patient-level health conditions) are close together. This step is crucial for incorporating the patient-specific factors, enabling the model to contextualize the cardiac health of each individual based on their unique physiological and demographic profile.

We apply average pooling to both imaging and tabular embeddings and use projectors to suppress irrelevant information, denoted $\mathcal{P}_I$ and $\mathcal{P}_T$, for the imaging and tabular embeddings, respectively. The final embeddings for image $Z_I$ and table $Z_T$ are thus given by:

$$Z_I = \mathcal{P}_I(\mathcal{E}_I(\tilde{X_I})), \quad Z_T = \mathcal{P}_T(\mathcal{E}_T(\tilde{X_T})). \tag{2}$$

Inspired by Li et al. (2023), we apply a 50% masking ratio to the imaging encoder during training. This strategy reduces the memory burden and allows for larger batch sizes during alignment, while preserving the effectiveness of the learned representation. No masking is applied to the tabular encoder.

Considering all subjects $\mathcal{N}$ in a training batch, the network is optimized with a loss function that pulls representations from the same subject closer while pushing representations from different subjects apart (Hager et al., 2023). The loss for this is defined as the negative similarity between image and tabular embeddings, decomposed into two parts: image-to-table loss and table-to-image loss. The image-to-table loss is formulated as:

$$\ell_{i,t} = -\sum_{j\in\mathcal{N}} \log \frac{\exp\left(\cos\left(z_{j_i}, z_{j_t}\right)/\tau\right)}{\sum_{k\in\mathcal{N},k\neq j}\exp\left(\cos\left(z_{j_i}, z_{k_t}\right)/\tau\right)}, \tag{3}$$

where $z_{j_i}$ and $z_{j_t}$ are the imaging embedding and tabular embedding of the $j$th subject in one batch, and $\tau$ is a small constant. The table-to-image loss $l_{t,i}$ is calculated analogously. With $\lambda$ as the loss weight, the total loss is thus

$$\mathcal{L} = \lambda\ell_{i,t} + (1-\lambda)\ell_{t,i}. \tag{4}$$

### 3.3. Stage III: Multi-task clinical evaluation with image encoder and task-specific decoders

In the final stage, we leverage the comprehensive cardiac representations learned in previous stages to enable a number of downstream, clinical tasks, including segmentation, phenotype/physiological feature prediction, and cardiac/metabolic disease classification, using aligned representations derived from CMR images. For each downstream task, the imaging encoder pretrained in Stage I and Stage II is fine-tuned jointly with a task-specific decoder. Both the encoder and decoder are fine-tuned separately for each task. These decoders are tailored to their respective clinical objectives, allowing the model to address diverse tasks within a unified and flexible framework.

It is important to note that all downstream tasks rely **solely on CMR images as input**, since the tabular data is too extensive to be feasibly collected during a standard clinical visit. Therefore, we performed all downstream tasks under the assumption that only CMR images are available.

For tabular data prediction, we employ a lightweight multi-head decoder with four specialized heads: a cardiac phenotype head, an anthropometric/physiological feature head, a binary classification head, and a multi-class classification head. For CMR sequence segmentation across all planes, we employ a U-Net-based decoder with skip connections (Zhou et al., 2023), ensuring accurate spatial localization and boundary refinement. For cardiac/metabolic disease classification with imbalanced classes, we adopt a lightweight task-specific decoder for each disease, allowing precise classification while maintaining computational efficiency. Note that the imaging encoder is not masked for any downstream tasks.

## 4. Experiments

### 4.1. Dataset

Our model is trained on paired CMR and tabular data from the UK Biobank (Petersen et al., 2015), using 42,000 subjects for pretraining the imaging encoder, aligning the imaging and tabular encoders, and fine-tuning on segmentation and patient-level health factor prediction. We derived disease classification labels from the “Diagnoses - ICD10-0” field in the UK Biobank (see Appendix F for details). Since only a subset of subjects have valid diagnostic entries, our disease classification task includes 38,000 subjects with confirmed labels for fine-tuning. We use the same 1000 subjects as a test set for all downstream task evaluations.

***CMR image data.*** The CMR sequence for each subject consists of 6 SA and 3 LA 2D slices, each with 50 time frames (see Appendix C for acquisition details). All slices are cropped to a size of 128 × 128 at the image center, max–min normalized to the range of 0 and 1, and augmented with random rotations, flipping, and contrast adjustments during training of all 3 stages. The segmentation labels are derived from images using the method proposed in Bai et al. (2020).

***Tabular data.*** We select in total 117 features, including anthropometric (i.e., sex, height, weight), clinical/physiological indicators (i.e., diabetes diagnosis, pulse rate), general lifestyle indicators (i.e., smoking frequency, medication intake), and cardiac SA-related phenotypes derived from segmentation labels (i.e., left ventricular mass, right ventricular ejection fraction) (see Appendix D for more details). LA-related phenotypes are only available for a few subjects, and thus, we do not include them in pretraining. Missing entries are imputed with the mean feature value for simplicity, maintaining the same sequence length. The numerical data fields were standardized using z-score normalization with a mean value of 0 and a standard deviation of 1.

### 4.2. Implementation details

***Stage I.*** We implement 6 encoder layers and 2 decoder layers with 5 attention heads each for imaging pretraining, using an embedding dimension of 1024. The imaging MAE employs a patch size of $8 \times 8 \times 5$, where $8 \times 8$ represents the spatial and 5 the temporal dimension. 5 frames are evenly selected from 50 and fixed for all subjects to reduce the computation cost. The masking ratio $q\%$ for the imaging encoder is set to 70% (more details about the ablation study are shown in Appendix A).

The batch size is set to 2 for imaging pretraining. The initial learning rate is set to $3\times10^{-3}$. We use a cosine annealing scheduler with warmup of 10 epochs and a weight decay of $10^{-4}$. The same scheduler is applied to all 3 stages. All training and test are conducted on a single NVIDIA A100 GPU.

***Stage II***. The tabular encoder is a transformer with two layers and 5 attention heads with an embedding dimension of 1025. 2 one-layer MLPs are applied after the encoders to project embeddings of both imaging and tabular data into a shared 128-dimensional space. The masking ratio $q\%$ for the imaging encoder is set to 50%.

In addition to masking the imaging encoder, we employ gradient checkpointing to reduce computational costs, enabling the use of a larger batch size (set to 256) for contrastive learning. This adjustment enhances the model's ability to distinguish and refine the learned representations. The initial learning rate is set to $3\times10^{-3}$. The temperature parameter $\tau$ used in the CLIP loss is 0.1, and the loss weight $\lambda$ in the CLIP loss is 0.5.

***Stage III***. The imaging encoder is pretrained through stage I and stage II and fully fine-tuned in stage III for different downstream tasks. For each task, the encoder is fine-tuned independently to optimize performance. The task-specific decoders are as follows: the segmentation decoder is U-Net-based (Zhou et al., 2023) with an embedding dimension of 576. The decoders of tabular reconstruction and disease classification have the same architecture, which is a two-layer MLP with a 256-dimensional embedding. The masking ratio $q\%$ for the imaging encoder is set to 0%.

The batch size is set to 2 for segmentation and 8 for tabular reconstruction and classification. The initial learning rate is set to $10^{-6}$ for downstream tasks.

***Baseline methods***. For tabular reconstruction, we use ResNet-50 (He et al., 2016), Vision Transformer (ViT) (Dosovitskiy et al., 2020), and an MAE model (Zhang et al., 2024a) as baselines. ResNet-50 serves as the competitive convolution-based baseline, representing an upper-bound benchmark for phenotype prediction. Since it performs single-target regression – unlike the proposed ViTa, which jointly predicts all targets – we assume the training set of 42,000 subjects provides sufficient scale to achieve state-of-the-art performance. The ViT baseline shares the same architecture as ViTa but is trained from scratch without any pretraining. It serves as the non-pretrained counterpart to ViTa, allowing us to isolate the effect of representation pretraining. The MAE baseline also adopts the same architecture as ViTa, consisting of a pretrained imaging encoder and a tabular reconstruction decoder. However, the imaging encoder in the MAE baseline is pretrained only on CMR data (Stage I) without the subsequent multi-modal alignment stage (Stage II). This design enables us to isolate and assess the specific contribution of tabular information in learning a more comprehensive and integrated cardiac representation.

For disease classification, we use ResNet-50 as the baseline model. For segmentation, we adopt the state-of-the-art nnU-Net (Isensee et al., 2021) as our upper bound, alongside single-view variants of our proposed ViTa model—ViTa (SA) and ViTa (LA)—which are pretrained and fine-tuned exclusively on short-axis (SA) or long-axis (LA) views, respectively.

## 5. Results and discussion

### 5.1. Integration of multi-modal information leads to coherent, information-enriched representations

We begin by visualizing the learned comprehensive, dense, and meaningful representations using t-SNE to showcase how the multi-view, multi-frame cardiac images and tabular data align effectively in a shared latent space.

In Fig. 4, each subplot is color-coded based on value groups of different cardiac phenotypes, such as LVEDV, LVSV, LVEF, LVCO, LVM, and other clinical indices listed in Appendix E. These phenotypes, derived from both LA and SA views, critically depend on spatio-temporal cardiac information. The observed clustering underscores the model's capability to effectively align spatio-temporal multi-view features, providing a rich, dense representation that encapsulates holistic cardiac information.

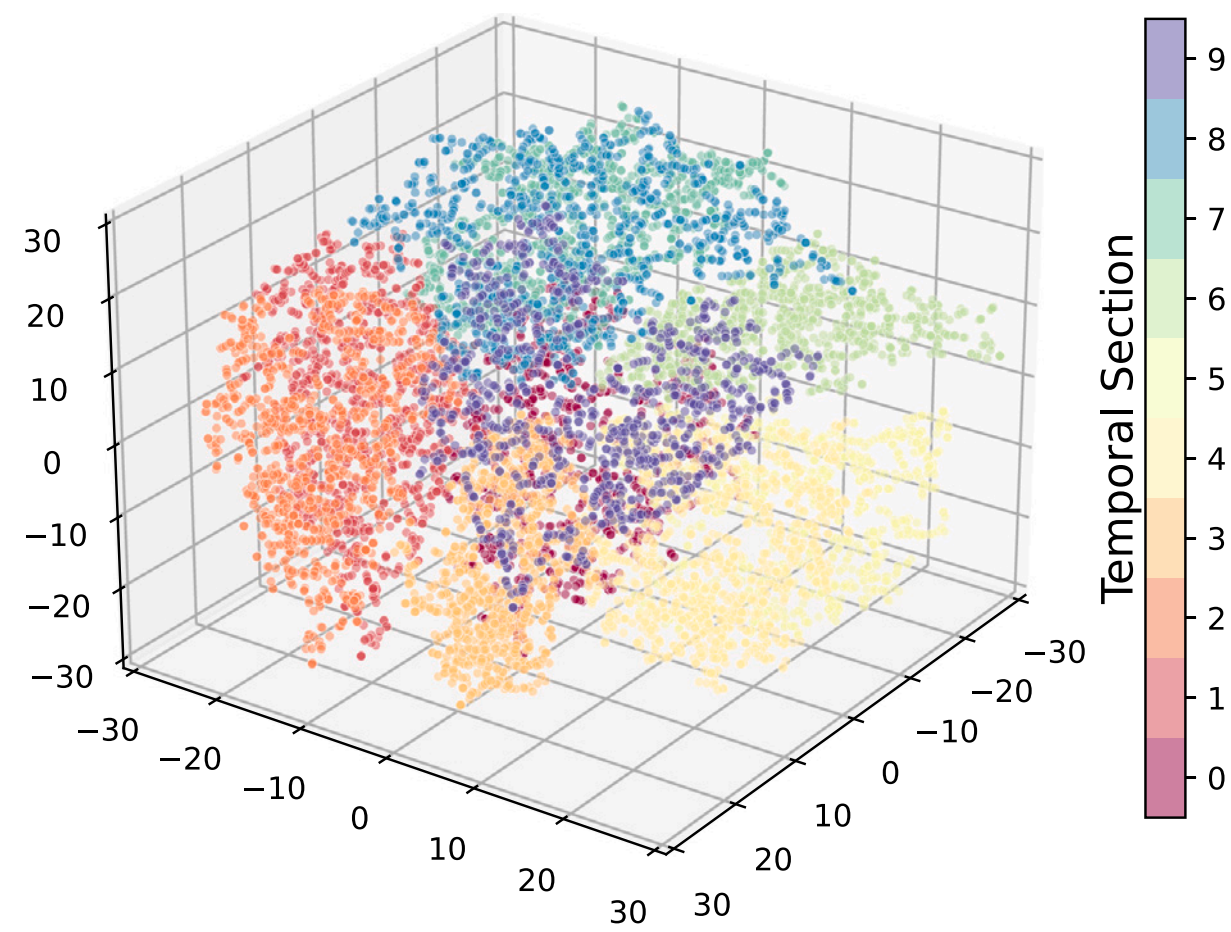

**Fig. 2.** t-SNE visualization of aligned cardiac representations from 1000 test subjects. Each subject is depicted using 10 temporal sections, each shown in a distinct color. Each temporal section consists of 5 cardiac phase frames, capturing the dynamic progression of cardiac motion.

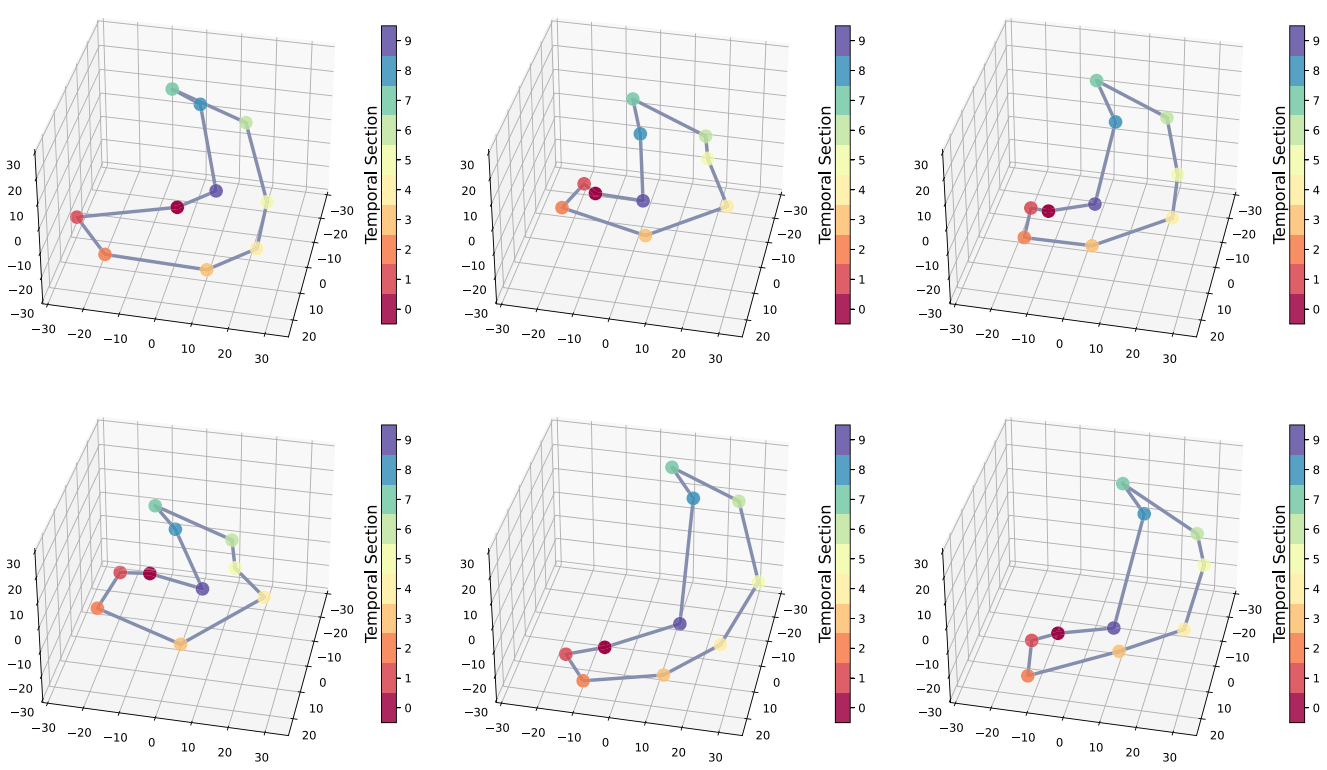

**Fig. 3.** t-SNE visualization showing temporal continuity of aligned cardiac representations at the subject level. Each subplot displays 10 temporal sections (5 sequential cardiac phase frames each) from one subject, connected in cardiac cycle order.

Furthermore, in Fig. 2, we illustrate the same representations with each subject represented by 10 distinct points, corresponding to 10 different time sections. Subject-level temporal continuity is further shown in Fig. 3, where temporal representations are connected in the cardiac cycle order. Notably, the temporal continuity is evident at the cohort level, with tight clustering within each time section, and an overlap between the first and last sections, reflecting the continuous nature of the cardiac motion. At the subject level, the temporal trajectory remains coherent, indicating preserved phase continuity. These observations underscore the model's spatio-temporal integration capability, demonstrating the alignment of temporal dynamics across multi-view CMR images.

These visualizations collectively emphasize the holistic integration of CMR and tabular data, establishing a strong foundation for the multi-task framework that follows in the subsequent stages of evaluation.

### 5.2. Performance in different downstream tasks

In this section, we show the results for SA/LA phenotype and physiological feature prediction, cardiac/metabolic disease classification with imbalanced classes, and all-plane segmentation, underscoring the generalizable efficiency of the learned comprehensive representations.

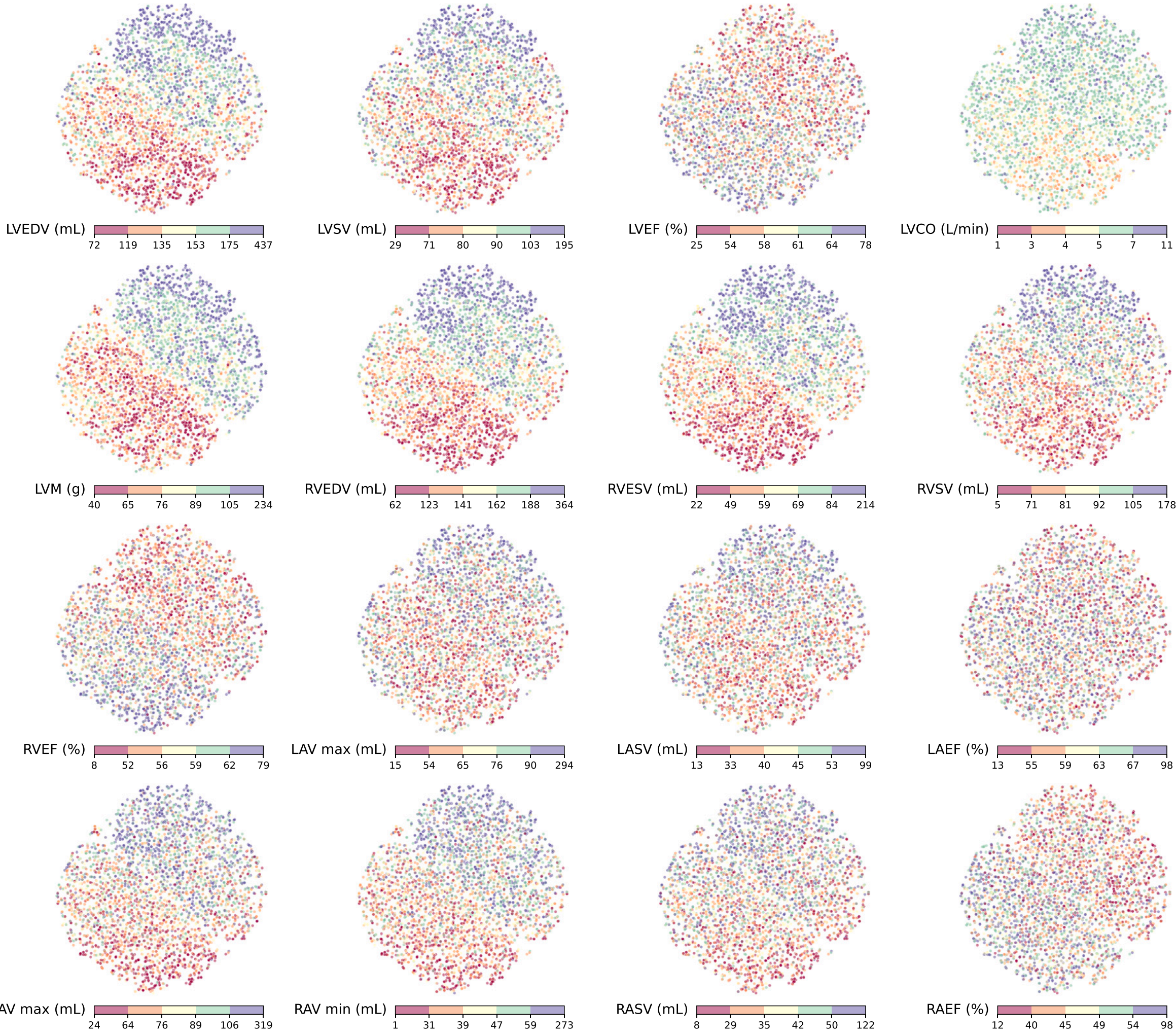

**Fig. 4.** The t-SNE visualization of the aligned cardiac representations from ViTa over 10,000 subjects. Subplots are color-coded by different phenotype groups.

#### 5.2.1. SA/LA phenotype and physiological/anthropometric feature prediction

***SA/LA phenotype prediction.*** Tables 1 and 2 present the mean absolute error for predicting SA and LA phenotypes **using solely CMR images** across several models: the mean-guess model (estimating each subject's phenotype value using the cohort mean), ResNet-50, ViT-based model (without any pretraining), MAE-based model (pretrained with only CMR images), and our proposed multi-modal ViTa. Given the large-scale training set comprising data from 42,000 subjects and the single-target regression setup, we adopt ResNet-50 as our convolution-based upper-bound baseline to leverage its strong feature extraction capabilities. It is important to note that our proposed model operates under a more challenging setting by predicting all phenotypes **simultaneously**, denoted as *all at once* in the results tables. In contrast, all baselines are trained to predict each phenotype **separately**, denoted as *individual*. Fig. 5 provides the regression plots and $R^2$ scores (the square of the sample correlation coefficient (r) between the true values and prediction) for each phenotype, illustrating predictions of 1000 subjects against the ground truth values.

ViTa achieves the smallest average mean absolute error across all SA phenotypes compared to all baselines, even though it tackles the more challenging task of predicting all phenotypes simultaneously. It underscores the richness of structural information embedded in the learned comprehensive cardiac representations. ViTa also surpasses the ViT-based model without any pretraining and the MAE-based model that is only pretrained with CMR data. It not only highlights the benefit of imaging pretraining and multi-modal alignment for cardiac holistic assessment, but also showcases how this alignment refines and shapes the learned representations to support more accurate phenotype-level interpretation.

Moreover, although LA phenotypes are excluded from the tabular data and thus never seen by the full framework during pretraining, ViTa still achieves comparable results to the convolution-based ResNet-50 and surpasses the ViT-based and MAE-based baselines. This demonstrates that the superior performance in phenotype prediction is not simply a result of integrating phenotype labels in stage II, but rather reflects ViTa's holistic understanding of cardiac anatomy and function.

Interestingly, we observe that *ViTa* shows less discriminative representations for LA-related features than SA, as evidenced by the less distinct clustering in Fig. 4 and lower $R^2$ score in Fig. 5. This is due to two factors: (1) limited LA-specific CMR views in the dataset and (2) the absence of LA phenotype values in tabular data during pretraining, leading the model to focus more on SA-specific features. As shown in Table 2, it struggles to predict LAV min, likely because ViTa predicts all LA phenotypes simultaneously, making it hard to balance different features.

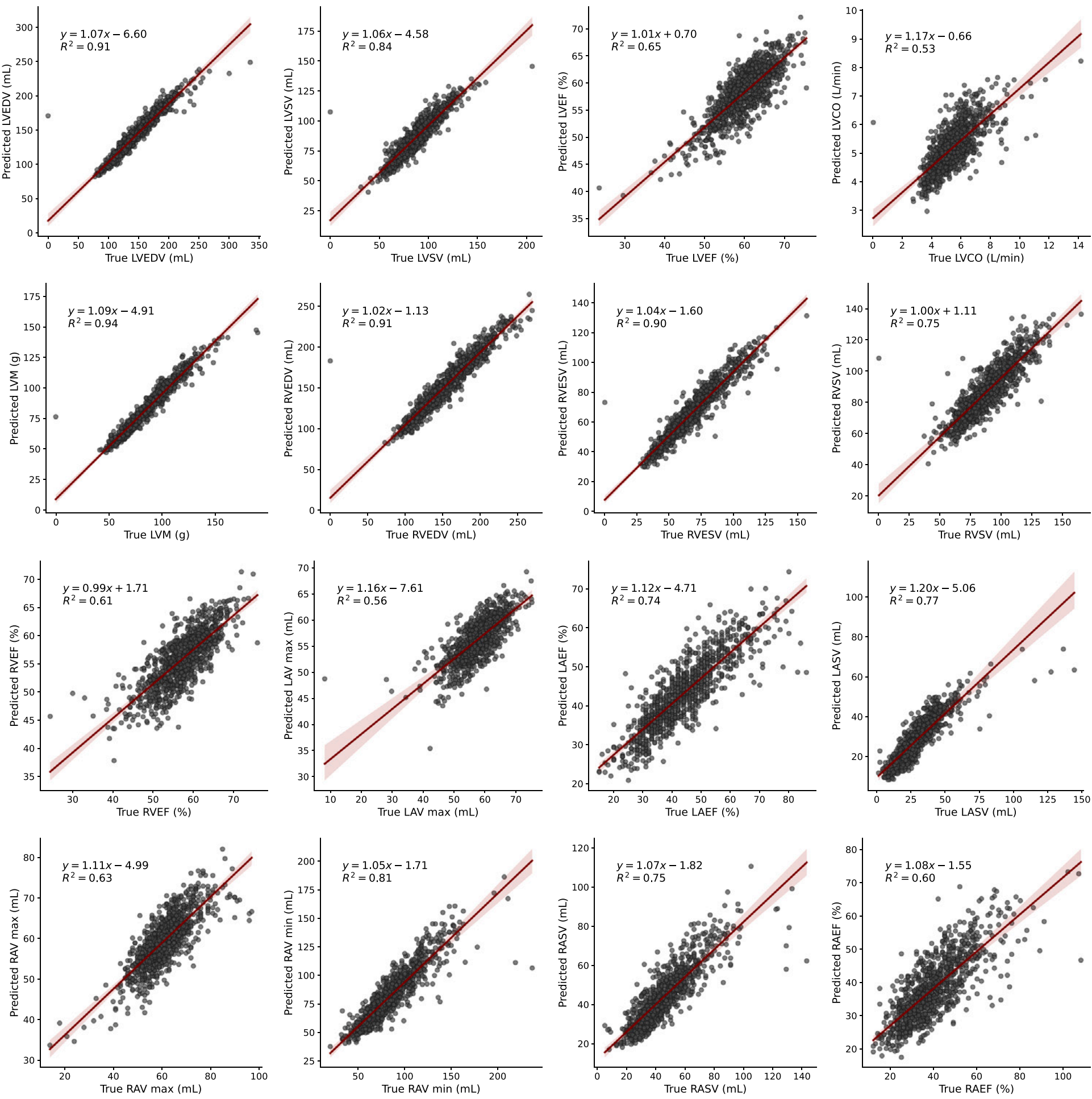

**Fig. 5.** Plots of ground truth LA and SA phenotypes against predictions from ViTa. The predictions are derived from CMR images of 1000 subjects. We present the regression line as well as the $R^2$ score for each phenotype.

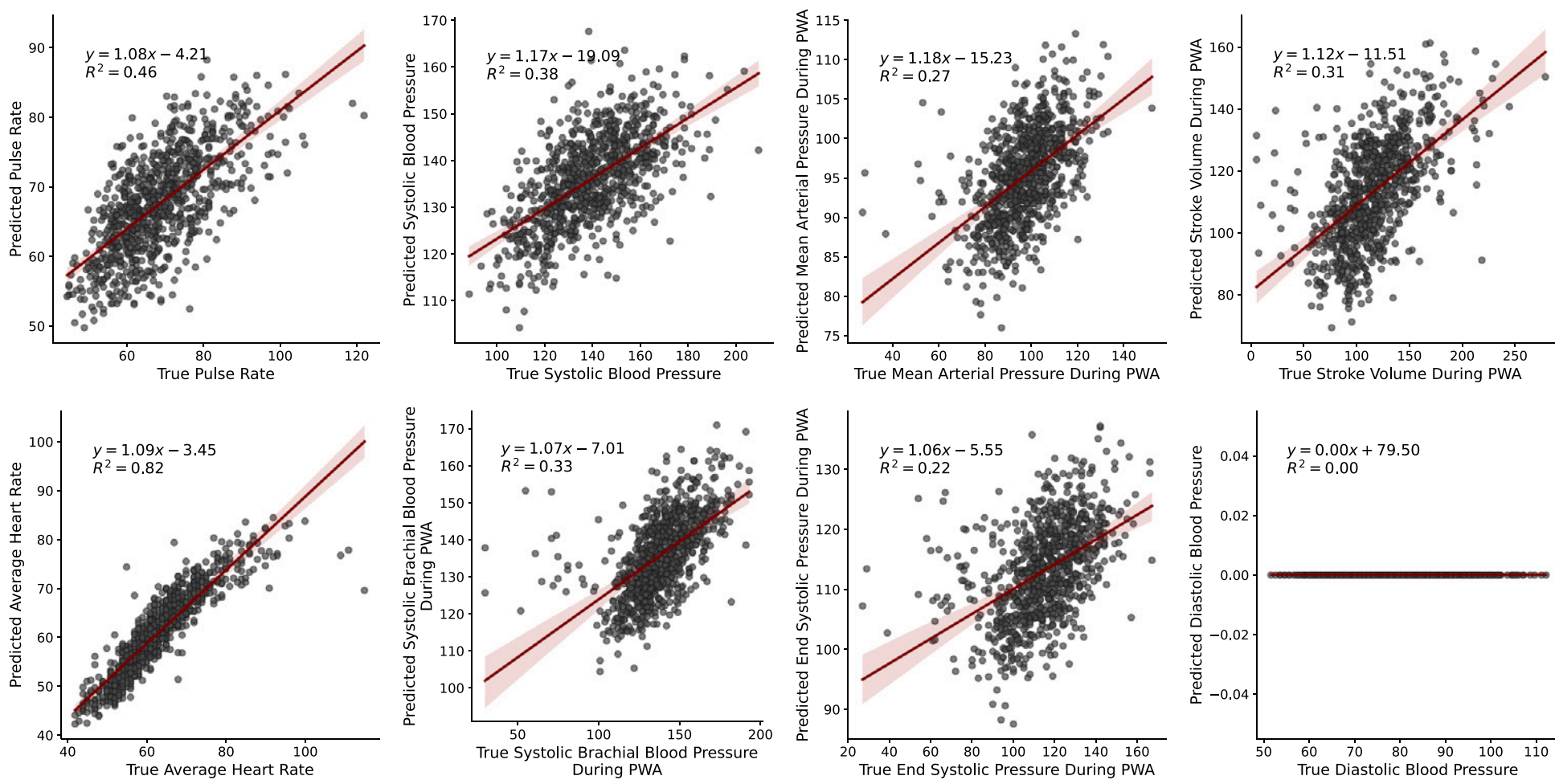

**Fig. 6.** Plots of ground truth physiological features against predictions from ViTa. The predictions are derived from CMR images of 1000 subjects. We present the regression line as well as the $R^2$ score for each feature.

**Table 1**
Comparison of mean absolute errors among mean-guess (estimating each subject's phenotype value using the cohort mean), ResNet-50, ViT, MAE-based model from Zhang et al. (2024a) using only images as input, and the proposed ViTa approach for SA phenotype prediction. Notably, only our proposed method predicts all phenotypes simultaneously rather than individually.

| Phenotype | Mean-guess | ResNet-50 (Individual) | ViT (Individual) | MAE (Individual) | ViTa (All at once) |
|---|---|---|---|---|---|
| LVEDV (mL) | $25.001_{\pm 20.393}$ | $\mathbf{5.398}_{\pm \mathbf{6.756}}$ | $11.526_{\pm 13.919}$ | $9.034_{\pm 10.474}$ | $6.193_{\pm 8.153}$ |
| LVSV (mL) | $14.442_{\pm 11.895}$ | $7.439_{\pm 8.404}$ | $6.930_{\pm 7.481}$ | $5.862_{\pm 5.815}$ | $\mathbf{5.190}_{\pm \mathbf{5.525}}$ |
| LVEF (%) | $4.571_{\pm 3.858}$ | $\mathbf{2.619}_{\pm \mathbf{2.196}}$ | $3.908_{\pm 3.079}$ | $3.199_{\pm 2.442}$ | $2.955_{\pm 2.438}$ |
| LVCO (L/min) | $0.957_{\pm 0.829}$ | $\mathbf{0.466}_{\pm \mathbf{0.474}}$ | $0.520_{\pm 0.543}$ | $0.490_{\pm 0.486}$ | $0.622_{\pm 0.653}$ |
| LVM (g) | $17.744_{\pm 12.880}$ | $\mathbf{3.967}_{\pm \mathbf{4.082}}$ | $7.848_{\pm 8.291}$ | $5.868_{\pm 5.390}$ | $4.146_{\pm 4.570}$ |
| RVEDV (mL) | $28.492_{\pm 20.565}$ | $8.705_{\pm 9.055}$ | $13.748_{\pm 13.973}$ | $9.493_{\pm 9.159}$ | $\mathbf{7.277}_{\pm \mathbf{8.081}}$** |
| RVESV (mL) | $16.395_{\pm 12.318}$ | $5.432_{\pm 5.251}$ | $8.404_{\pm 7.944}$ | $6.066_{\pm 6.561}$ | $\mathbf{4.693}_{\pm \mathbf{4.752}}$** |
| RVSV (mL) | $15.298_{\pm 11.566}$ | $7.439_{\pm 8.404}$ | $9.506_{\pm 8.498}$ | $7.419_{\pm 6.904}$ | $\mathbf{6.839}_{\pm \mathbf{6.658}}$** |
| RVEF (%) | $4.801_{\pm 3.861}$ | $\mathbf{3.038}_{\pm \mathbf{2.611}}$ | $4.168_{\pm 3.366}$ | $3.616_{\pm 3.021}$ | $3.225_{\pm 2.555}$ |
| Average | $14.189_{\pm 10.907}$ | $4.945_{\pm 5.248}$ | $7.197_{\pm 6.836}$ | $5.672_{\pm 5.584}$ | $\mathbf{4.571}_{\pm \mathbf{4.821}}$ |

** Statistical significance at $p < 0.01$ compared to the best-performing baseline.

**Table 2**
Comparison of mean absolute errors among mean-guess (estimating each subject's phenotype value using the cohort mean), ResNet-50, ViT, MAE-based model from Zhang et al. (2024a) using only images as input, and the proposed ViTa approach for LA phenotype prediction. Notably, only our proposed method predicts all phenotypes simultaneously rather than individually.

| Phenotype | Mean-guess | ResNet-50 (Individual) | ViT (Individual) | MAE (Individual) | ViTa (All at once) |
|---|---|---|---|---|---|
| LAV max (mL) | $4.662_{\pm 4.066}$ | $5.602_{\pm 6.215}$ | $8.889_{\pm 7.704}$ | $6.115_{\pm 5.972}$ | $\mathbf{3.270}_{\pm \mathbf{2.882}}$** |
| LAV min (mL) | $17.649_{\pm 14.314}$ | $3.173_{\pm 3.398}$ | $4.565_{\pm 4.789}$ | $3.964_{\pm 4.105}$ | n.a. |
| LASV (mL) | $10.277_{\pm 10.132}$ | $4.479_{\pm 3.970}$ | $5.554_{\pm 4.766}$ | $\mathbf{4.103}_{\pm \mathbf{3.724}}$ | $4.524_{\pm 5.692}$ |
| LAEF (%) | $9.366_{\pm 7.280}$ | $\mathbf{4.073}_{\pm \mathbf{3.909}}$ | $5.238_{\pm 4.368}$ | $4.290_{\pm 3.638}$ | $4.562_{\pm 4.222}$ |
| RAV max (mL) | $6.702_{\pm 6.115}$ | $6.348_{\pm 7.895}$ | $11.017_{\pm 11.563}$ | $7.535_{\pm 8.160}$ | $\mathbf{4.243}_{\pm \mathbf{3.901}}$** |
| RAV min (mL) | $20.532_{\pm 17.188}$ | $\mathbf{4.270}_{\pm \mathbf{5.441}}$ | $7.881_{\pm 7.930}$ | $4.754_{\pm 5.577}$ | $7.971_{\pm 8.722}$ |
| RASV (mL) | $13.586_{\pm 11.593}$ | $5.500_{\pm 4.870}$ | $7.438_{\pm 6.713}$ | $\mathbf{5.469}_{\pm \mathbf{5.086}}$ | $6.272_{\pm 6.533}$ |
| RAEF (%) | $10.124_{\pm 8.491}$ | $\mathbf{4.867}_{\pm \mathbf{4.066}}$ | $6.092_{\pm 5.011}$ | $5.047_{\pm 4.101}$ | $6.159_{\pm 5.850}$ |
| Average | $10.750_{\pm 9.266}$ | $\mathbf{5.020}_{\pm \mathbf{5.195}}$ | $7.084_{\pm 6.605}$ | $5.330_{\pm 5.180}$ | $5.286_{\pm 5.400}$ |

** Statistical significance at $p < 0.01$ compared to the best-performing baseline.

**Table 3**
Comparison of mean absolute errors between mean-guess (estimating each subject's measurement using the cohort mean), ResNet-50, and ViTa for various cardiac-related physiological features and anthropometric features.

| Measurement | Mean-guess | ResNet-50 (Individual) | ViTa (All at once) |
|---|---|---|---|
| Systolic blood pressure during PWA (mmHg) | $14.782_{\pm 10.839}$ | $12.938_{\pm 9.889}$ | $\mathbf{11.638}_{\pm \mathbf{9.375}}$** |
| Pulse rate (*bpm*) | $9.053_{\pm 7.093}$ | $7.141_{\pm 6.010}$ | $\mathbf{6.619}_{\pm \mathbf{5.424}}$** |
| Mean arterial pressure during PWA (mmHg) | $10.197_{\pm 8.701}$ | $9.012_{\pm 8.184}$ | $\mathbf{8.643}_{\pm \mathbf{7.807}}$** |
| Stroke volume during PWA (mL) | $25.488_{\pm 22.238}$ | $23.617_{\pm 21.529}$ | $\mathbf{20.545}_{\pm \mathbf{19.364}}$** |
| Systolic brachial blood pressure during PWA (mmHg) | $14.517_{\pm 13.167}$ | $12.635_{\pm 12.586}$ | $\mathbf{11.378}_{\pm \mathbf{11.557}}$** |
| Average heart rate (*bpm*) | $7.745_{\pm 6.566}$ | $\mathbf{3.271}_{\pm \mathbf{3.079}}$ | $3.331_{\pm 3.501}$ |
| End systolic pressure during PWA (mmHg) | $13.623_{\pm 11.858}$ | $12.517_{\pm 11.710}$ | $\mathbf{11.711}_{\pm \mathbf{10.989}}$** |
| Age (*years*) | $6.203_{\pm 4.261}$ | $4.083_{\pm 3.249}$ | $\mathbf{3.738}_{\pm \mathbf{2.752}}$** |
| Body fat percentage (%) | $6.338_{\pm 4.537}$ | $\mathbf{2.749}_{\pm \mathbf{2.324}}$ | $3.015_{\pm 2.294}$ |
| Body mass index (kg/m$^2$) | $3.186_{\pm 2.702}$ | $1.525_{\pm 1.413}$ | $\mathbf{1.518}_{\pm \mathbf{1.324}}$** |
| Height (cm) | $7.681_{\pm 5.153}$ | $\mathbf{4.562}_{\pm \mathbf{3.668}}$ | $5.743_{\pm 4.271}$ |
| Weight (kg) | $11.931_{\pm 8.895}$ | $6.112_{\pm 5.185}$ | $\mathbf{4.216}_{\pm \mathbf{3.677}}$** |
| **Average** | $10.868_{\pm 7.724}$ | $8.347_{\pm 7.402}$ | $\mathbf{7.675}_{\pm \mathbf{6.861}}$ |

** Statistical significance at $p < 0.01$ compared to the best-performing baseline.

***Physiological/anthropometric feature prediction.*** Apart from the cardiac phenotypes, we also show our model's physiological/anthropometric prediction ability, comparing to ResNet-50, which is the most competitive baseline shown in the previous section. Results are presented in Fig. 6 and Table 3. The superior performance of ViTa over the baseline underscores its effectiveness for patient-specific health assessment.

### 5.2.2. Cardiac and metabolic disease classification under class imbalance

We also highlight the robustness of the model in handling cardiac and metabolic disease classification with imbalanced classes.

As shown in Table 4, our multi-modal model achieves superior performance across all 5 cardiac and metabolic disease classification tasks, significantly outperforming ResNet-50, which relies solely on imaging data. The ResNet-50 baseline tends to bias toward the majority class, often failing to correctly identify true positive cases, which is a critical limitation in real-world clinical applications.

By integrating image-derived structural information with patient-level health factors, our approach better captures disease heterogeneity. This highlights the importance of multi-modal fusion, where combining CMR and tabular factors provides a more patient-specific and comprehensive assessment of cardiac health. These findings reinforce the potential of our model as a general-purpose foundation for multi-task cardiac assessment, a crucial step toward more personalized and data-driven cardiovascular healthcare.

**Table 4**
Comparison of classification performance between the image-based ResNet-50 and our image-tabular ViTa model across six cardiac and metabolic diseases. Metrics include AUC-ROC, F1 Score, Recall, Precision, and Average Precision (AP). The positive class ratio for each disease is shown as a percentage of the total subject population.

| Disease | Method | AUC-ROC ↑ | F1 Score ↑ | Recall ↑ | Precision ↑ | AP ↑ |
|---|---|---|---|---|---|---|
| CAD | ResNet-50 | 0.608 | 0 | 0 | 0 | 0.139 |
| (7.4%) | ViTa (ours) | **0.770** | **0.312** | **0.233** | **0.472** | **0.333** |
| Infarct | ResNet-50 | 0.655 | 0 | 0 | 0 | 0.096 |
| (2.9%) | ViTa (ours) | **0.817** | **0.318** | **0.241** | **0.467** | **0.260** |
| High blood pressure | ResNet-50 | 0.685 | 0.095 | 0.054 | 0.359 | 0.397 |
| (25.8%) | ViTa (ours) | **0.778** | **0.531** | **0.595** | **0.480** | **0.507** |
| Diabetes | ResNet-50 | 0.578 | 0 | 0 | 0 | 0.061 |
| (4.4%) | ViTa (ours) | **0.813** | **0.111** | **0.070** | **0.273** | **0.218** |
| Hypertension | ResNet-50 | 0.607 | 0.047 | 0.024 | **0.714** | 0.296 |
| (20.8%) | ViTa (ours) | **0.746** | **0.461** | **0.517** | 0.416 | **0.428** |

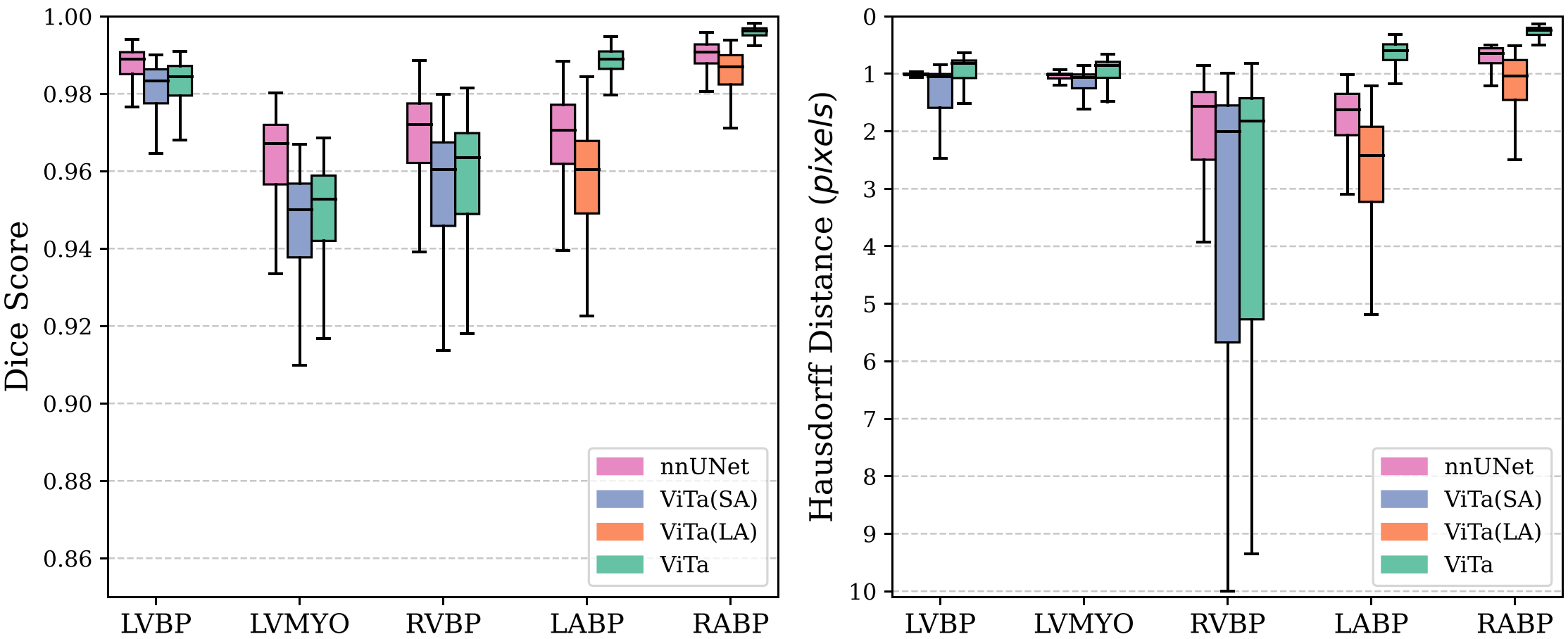

**Fig. 7.** Segmentation performance evaluated using Dice score and Hausdorff Distance (HD) across 1000 subjects. Results are shown for nnUNet, ViTa (SA) (pretrained and fine-tuned using only short-axis slices), ViTa (LA) (using only long-axis slices), and ViTa (using both short-axis and long-axis slices for pretraining and fine-tuning). Segmentation targets include the left ventricular blood pool (LVBP), left ventricular myocardium (LVMYO), right ventricular blood pool (RVBP), and right atrial blood pool (RABP).

*5.2.3. All-plane all-view segmentation*

Our model demonstrates promising segmentation accuracy using only imaging features. As shown in Figs. 7 and 8, our model achieves strong qualitative and quantitative results in all-plane, multi-view, and multi-frame segmentation, highlighting the model's ability to accurately delineate cardiac structures. The strong performance without tabular input suggests that the multi-view, multi-frame CMR data provides sufficient anatomical information for segmentation.

Interestingly, while multi-modal fusion improves phenotype regression and disease classification, we observe that segmentation relies predominantly on the rich spatio-temporal information present in CMR images. This suggests that holistic cardiac assessment benefits from task-specific optimization, where structural tasks (e.g., segmentation) depend primarily on visual data, whereas functional and diagnostic tasks (e.g., phenotype prediction, disease classification) gain from integrating systemic health factors.

*5.3. Limitation and outlook*

While *ViTa* demonstrates strong performance across multiple cardiac assessment tasks, we have not explored tasks such as motion estimation/mesh prediction for a more comprehensive evaluation of the learned cardiac representation on cardiac health assessment. In future work, we aim to extend this framework to a broader range of tasks and incorporate additional modalities, enhancing the depth and accuracy of cardiac representations. Furthermore, we plan to explore the association between patient health factors – such as cardiac phenotypes, anthropometric data, and lifestyle factors – and cardiac health in a correlation study. Additionally, incorporating genetic data will further enable a more comprehensive, patient-specific cardiac assessment, providing valuable insights into the underlying factors influencing cardiac health.

## 6. Conclusion

In this work, we introduce *ViTa*, the first fully multi-view, multi-modal, and multi-task framework for comprehensive cardiac function assessment, integrating image-derived, phenotypic, and physiological data to generate patient-specific cardiac representations. We demonstrate that these cardiac holistic representations serve as a strong foundation and enable multi-task learning across phenotype and physiological prediction, cardiac and metabolic disease classification under class imbalance, and segmentation. By bridging the gap between imaging and patient-level health factors, our work represents a crucial step toward a foundational cardiac health assessment framework, paving the way for more accurate, personalized, and scalable cardiovascular diagnostics.

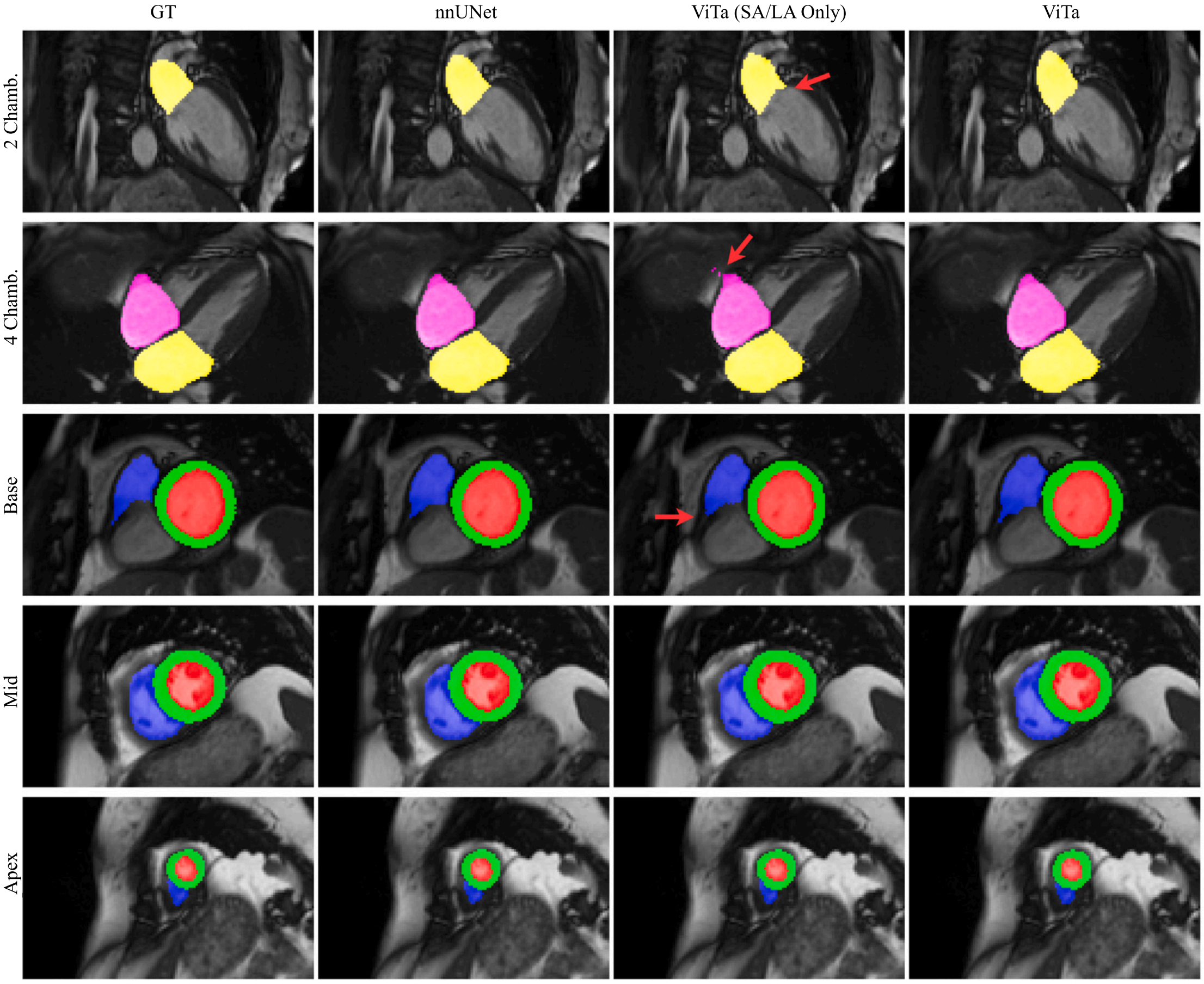

**Fig. 8.** Examples of predicted segmentation maps for the 2-chamber and 4-chamber long-axis views, as well as base, mid, and apex short-axis views. Results are shown for nnUNet, ViTa (SA/LA only) (pretrained and fine-tuned using only short-axis or long-axis slices), and ViTa (pretrained and fine-tuned using both short-axis and long-axis slices).

## 7. Description of the extensions

This paper extends our MICCAI 2024 main conference paper (Zhang et al., 2024a). The following extension is made based on our original paper: (1) We extend the methodology by incorporating a broad range of tabular features, including phenotypes, anthropometric features, physiological features, and lifestyle data, alongside 3D+T CMR sequences. This ensures more interpretable and patient-specific representations for cardiac health assessment. (2) We perform a latent space analysis on spatially sensitive features, including both SA and LA phenotypes, and further examine the temporal dynamics of the latent space in relation to the cardiac cycle, whereas in our previous paper, we only focused on a very limited set of phenotypes. (3) We carry out more comprehensive experiments, evaluating the efficiency of representations as the foundation of cardiac assessment by performing multi-task, including a wide range of SA/LA phenotype/physiological feature simultaneous prediction, segmentation, and cardiac and metabolic disease classification. In contrast, our previous work focused only on separate SA phenotype prediction and segmentation.

## CRediT authorship contribution statement

**Yundi Zhang:** Writing – review & editing, Writing – original draft, Visualization, Validation, Supervision, Software, Resources, Project administration, Methodology, Investigation, Formal analysis, Data curation, Conceptualization. **Paul Hager:** Writing – review & editing. **Che Liu:** Methodology. **Suprosanna Shit:** Conceptualization. **Chen Chen:** Conceptualization. **Daniel Rueckert:** Supervision, Methodology, Funding acquisition. **Jiazhen Pan:** Writing – review & editing, Supervision, Methodology, Conceptualization.

## Declaration of competing interest

The authors declare that they have no known competing financial interests or personal relationships that could have appeared to influence the work reported in this paper.

## Acknowledgments

This research has been conducted using the UK Biobank Resource under Application Number 87 802. This work is funded by the European Research Council (ERC) project Deep4MI (884622).

## Appendix A. Ablation study

We performed an ablation study evaluating the impact of varying masking ratios and the inclusion of 5D positional encoding on model pretraining performance. The qualitative and quantitative results are presented in Tables A.5, A.6, and Fig. B.9. Our findings emphasize the importance of including an additional dimension to represent different views for effective multi-view reconstruction. Furthermore, considering that contrastive learning in stage II is highly sensitive to batch size, we selected a masking ratio of 0.7 to balance reconstruction quality and computational efficiency.

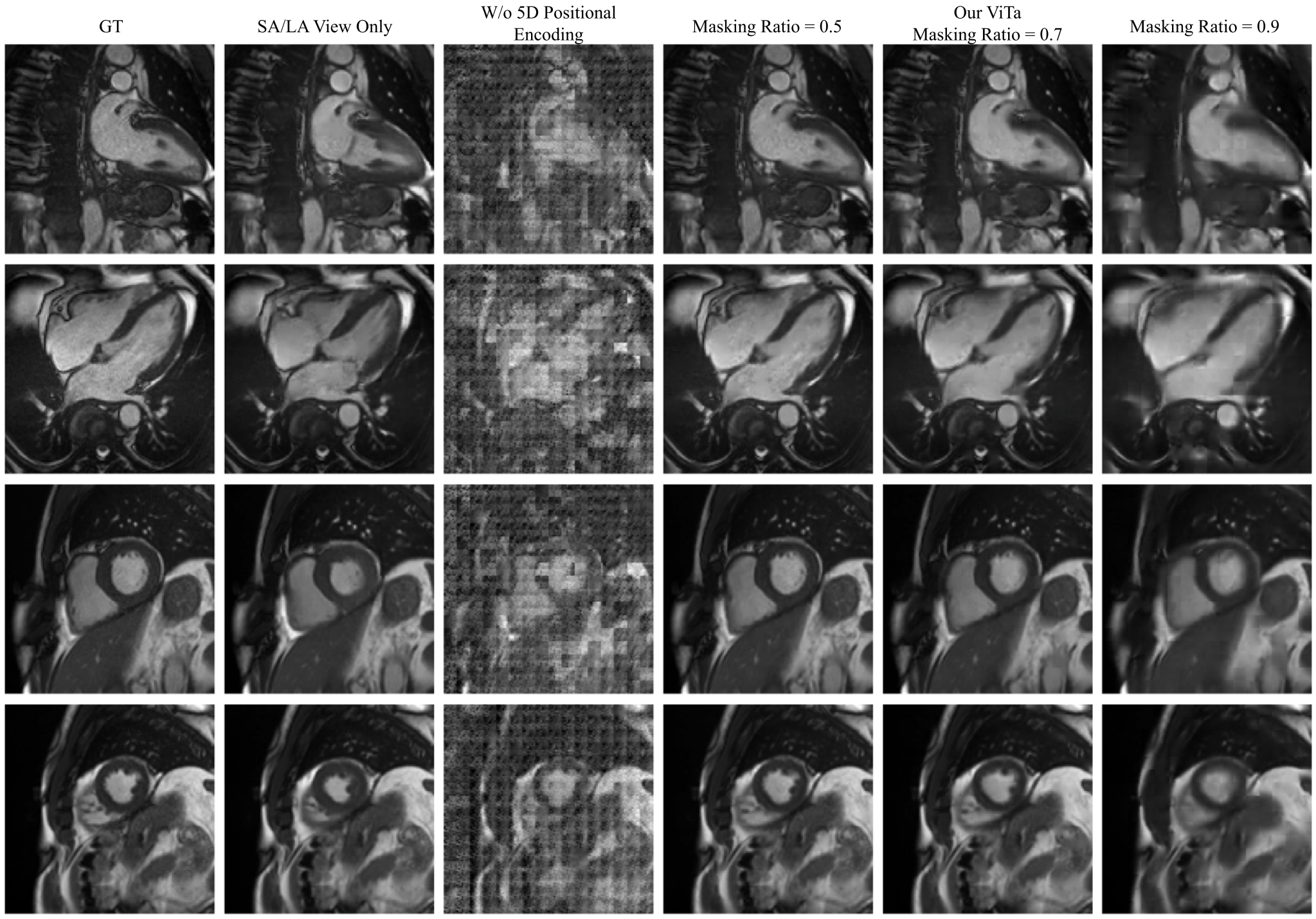

**Fig. B.9.** Reconstruction examples during pre-training. The second column shows results using input slices from a single view (SA/LA only), while the remaining columns use both SA and LA views. The third column corresponds to a model without 5D positional encoding. The last three columns present reconstructions with different masking ratios: 0.5, 0.7 (our ViTa), and 0.9.

**Table A.5**
Comparison of PSNR for pretraining reconstruction using models with only 4D positional encoding (PE) and with 5D PE including an additional dimension indicating different views.

| Model | PSNR (dB) |
|---|---|
| 4D PE | $16.78_{\pm 0.90}$ |
| 5D PE | $33.59_{\pm 2.19}$ |

**Table A.6**
Comparison of PSNR for pretraining reconstruction under varying masking ratios.

| Masking ratio | PSNR (dB) |
|---|---|
| 0.5 | $38.03_{\pm 2.56}$ |
| 0.7 | $33.59_{\pm 2.19}$ |
| 0.9 | $27.21_{\pm 1.91}$ |

## Appendix B. Positional embeddings

The positional embeddings consist of $PE_{SA}$, $PE_{LA}$, and view indicator $PE_{indicator}$. The positional embeddings of SA can be formulated as:

$$PE_{SA}^{(pos,2i)} = \sin\left(pos/10000^{2i/(dim-1)}\right)$$
$$PE_{SA}^{(pos,2i+1)} = \cos\left(pos/10000^{2i/(dim-1)}\right), \tag{B.1}$$

where $pos$ is the position of the SA grid $(x - y - t - z)$ with $z$ as the number of SA slices. $PE_{LA}$ is calculated analogously. The view indicator $PE_{indicator} \in \mathcal{R}^{L\times 1}$ is then concatenated to the stack of $PE_{SA}$ and $PE_{LA}$ to form positional embeddings of shape $(L \times dim)$.

## Appendix C. Data acquisition

We report the acquisition parameters of CMR images for long-axis and short-axis views, including slice thickness, slice gap, and voxel size, in Table C.7. Additionally, cohort-level means and standard deviations for all physiological features, as well as short-axis (SA) and long-axis (LA) phenotype features of the test dataset, are provided in Table C.8.

**Table C.7**
Imaging data acquisition parameters for long-axis and short-axis views.

| Description | LA | SA |
|---|---|---|
| Slice thickness (mm) | 6.0 | 8.0 |
| Slice gap (mm) | N.A. | 2 |
| Voxel size | 1.8 × 1.8 × 6.0 | 1.8 × 1.8 × 8.0 |

## Appendix D. Tabular features selection

The features we select for tabular input are as follows:

- Numerical features: Pulse wave Arterial Stiffness index, Systolic blood pressure (mean), Diastolic blood pressure (mean), Pulse rate (mean), Body fat percentage, Whole body fat mass, Whole body fat-free mass, Whole body water mass, Body mass index (BMI), Cooked vegetable intake, Salad/raw vegetable intake, Cardiac operations performed, Total mass, Basal metabolic rate, Impedance of whole body, Waist circumference, Hip circumference, Standing height, Height, Sitting height, Weight, Ventricular rate, P duration, QRS duration, PQ interval, RR interval, PP interval, Cardiac output, Cardiac index, Average heart rate, Body surface area, Duration of walks, Duration of moderate activity, Duration of vigorous activity, Time spent watching television (TV), Time spent using computer, Time spent driving, Heart rate during PWA, Systolic brachial blood pressure during PWA, Diastolic brachial blood pressure during PWA, Peripheral pulse pressure during PWA, Central systolic blood pressure during PWA, Central pulse pressure during PWA, Number of beats in waveform average for PWA, Central augmentation pressure during PWA, Augmentation index for PWA, Cardiac output during PWA, End systolic pressure during PWA, End systolic pressure index during PWA, Stroke volume during PWA, Mean arterial pressure during PWA, Cardiac index during PWA, Sleep duration, Exposure to tobacco smoke at home, Exposure to tobacco smoke outside home, Pack years of smoking, Pack years adult smoking as proportion of life span exposed to smoking, LVEDV (mL), LVESV (mL), LVSV (mL), LVEF (%), LVCO (L/min), LVM (g), RVEDV (mL), RVESV (mL), RVSV (mL), and RVEF (%).

**Table C.8**
Mean and standard deviation using cohort-level mean estimates for all physiological and anthropometric features, as well as short-axis (SA) and long-axis (LA) phenotype features.

| Physiological& Anthropometric | Systolic BP | Pulse rate | MAP | Stroke vol. | Brachial BP | HR avg | ESP | Age | Body fat | BMI | Height | Weight |
|---|---|---|---|---|---|---|---|---|---|---|---|---|
| Mean (MAE) | 14.782 | 9.053 | 10.197 | 25.488 | 14.517 | 7.745 | 13.623 | 6.203 | 6.338 | 3.186 | 7.681 | 11.931 |
| Std | 10.839 | 7.093 | 8.701 | 22.238 | 13.167 | 6.566 | 11.858 | 4.261 | 4.537 | 2.702 | 5.153 | 8.895 |

| SA phenotypes | LVEDV | LVSV | LVEF | LVCO | LVM | RVEDV | RVESV | RVSV | RVEF |
|---|---|---|---|---|---|---|---|---|---|
| Mean | 25.001 | 14.442 | 4.571 | 0.957 | 17.744 | 28.492 | 16.395 | 15.298 | 4.801 |
| Std | 20.393 | 11.895 | 3.858 | 0.829 | 12.880 | 20.565 | 12.318 | 11.566 | 3.861 |

| LA phenotypes | LAV max | LAV min | LASV | LAEF | RAV max | RAV min | RASV | RAEF |
|---|---|---|---|---|---|---|---|---|
| Mean | 4.662 | 17.649 | 10.277 | 9.366 | 6.702 | 20.532 | 13.586 | 10.124 |
| Std | 4.066 | 14.314 | 10.132 | 7.280 | 6.115 | 17.188 | 11.593 | 8.491 |

- Binary categorical features: Worrier/anxious feelings, Shortness of breath walking on level ground, Sex, Diabetes diagnosis, Heart attack diagnosed by doctor, Angina diagnosed by doctor, Stroke diagnosed by doctor, High blood pressure diagnosed by doctor, Cholesterol lowering medication regularly taken, Blood pressure medication regularly taken, Insulin medication regularly taken, Hormone replacement therapy medication regularly taken, Oral contraceptive pill or minipill medication regularly taken, Pacemaker, Ever had diabetes (Type I or Type II), Long-standing illness, disability or infirmity, Tense/'highly strung', and Ever smoked.
- Sleeplessness/insomnia, Frequency of heavy DIY in last 4 weeks, Alcohol intake frequency, Processed meat intake, Beef intake, Pork intake, Lamb/mutton intake, Overall health rating, Alcohol usually taken with meals, Alcohol drinker status, Frequency of drinking alcohol, Frequency of consuming six or more units of alcohol, Amount of alcohol drunk on a typical drinking day, Falls in the last year, Weight change compared with 1 year ago, Number of days/week walked 10+ min, Number of days/week of moderate physical activity 10+ min, Number of days/week of vigorous physical activity 10+ min, Usual walking pace, Frequency of stair climbing in last 4 weeks, Frequency of walking for pleasure in last 4 weeks, Duration walking for pleasure, Frequency of strenuous sports in last 4 weeks, Duration of strenuous sports, Duration of light DIY, Duration of heavy DIY, Frequency of other exercises in last 4 weeks, Duration of other exercises, Current tobacco smoking, Past tobacco smoking, Smoking/smokers in household, and Smoking status.

## Appendix E. Cardiac phenotypes

The cardiac phenotypes shown in the work include Left Ventricular End-Diastolic Volume (LVEDV), Left Ventricular Stroke Volume (LVSV), Left Ventricular Ejection Fraction (LVEF), Left Ventricular Cardiac Output (LVCO), Left Ventricular Mass (LVM), Right Ventricular End-Diastolic Volume (RVEDV), Right Ventricular End-Systolic Volume (RVESV), Right Ventricular Stroke Volume (RVSV), Right Ventricular Ejection Fraction (RVEF), Left Atrium Volume Maximum (LAV max), Left Atrium Volume Minimum (LAV min), Left Atrium Stroke Volume (LASV), Left Atrium Ejection Fraction (LAEF), Right Atrium Volume Maximum (RAV max), Right Atrium Volume Minimum (RAV min), Right Atrium Stroke Volume (RASV), and Right Atrium Ejection Fraction (RAEF).

## Appendix F. Classification labels

Based on Diagnoses - ICD10-0 of UK BioBank, we label the patients based on the following entries:

- CAD (coronary artery disease): I200, I201, I208, I209, I220, I221, I228, I229, I210, I211, I212, I213, I214, I219, I240, I248, I249, I250, I251, I252, I253, I254, I255, I256, I258, I259.
- Stroke: I630, I631, I632, I633, I634, I635, I636, I638, I639.
- Hypertension: I10, I110, I119, I120, I129, I130, I131, I132, I139, I150, I151, I152, I158, I159.
- Infarct: I210, I211, I212, I213, I214, I219, I252.
- Diabetes: E100, E101, E102, E103, E104, E105, E106, E107, E108, E109, E110, E111, E112, E113, E114, E115, E116, E117, E118, E119, E121, E123, E125, E128, E129, E130, E131, E132, E133, E134, E135, E136, E137, E138, E139, E140, E141, E142, E143, E144, E145, E146, E147, E148, E149.